\documentclass[showpacs,amsmath,amssymb]{revtex4}

\usepackage{bm}

\begin{document}

\newcommand{\uu}[1]{\underline{#1}}
\newcommand{\pp}[1]{\phantom{#1}}
\newcommand{\be}{\begin{eqnarray}}
\newcommand{\ee}{\end{eqnarray}}
\newcommand{\ve}{\varepsilon}
\newcommand{\vs}{\varsigma}
\newcommand{\Tr}{{\,\rm Tr\,}}
\newcommand{\pol}{\textstyle{\frac{1}{2}}}

\title{
Regularization as quantization in reducible representations of CCR
}
\author{Marek Czachor$^{1,2}$ and Jan Naudts$^2$}
\affiliation{
$^1$ Katedra Fizyki Teoretycznej i Metod Matematycznych\\
Politechnika Gda\'nska, 80-952 Gda\'nsk, Poland\\
$^2$ Departement Fysica, Universiteit Antwerpen, B2610 Antwerpen, Belgium\\
}

\begin{abstract}
A covariant quantization scheme employing reducible representations of
canonical commutation relations with positive-definite metric and
Hermitian four-potentials (an alternative to the Gupta-Bleuler method)
is tested on the example of quantum electromagnetic fields produced
by a classical current. The Heisenberg dynamics can be consistently
formulated since the fields are given by operators and not operator-valued
distributions.
The scheme involves a Hamiltonian whose free part is modified but the
minimal-coupling interaction is the standard one. Solving Heisenberg equations
of motion under the assumption that the fields are free for times $t_0=\pm\infty$
we arrive at retarded and advanced solutions. Once we have these solutions we can
deduce the form of evolution of retarded and advanced fields between two arbitrary
{\it finite\/} times. The appropriate unitary evolution operators are found and
their generators are computed. Now the generators involve the same free part as
before, but the interaction term turns out to be modified. For a pointlike charge
localized on a world-line $z^a(t)$ we find the interaction term of the form
$-q\vec{A}\big(z(t)\big)\cdot\vec v(t)
-q\int d\vec z\cdot\vec {E}$
where the integration is along those parts of the charge world-line where the
charge velocity is nonzero. There is no self-energy contribution.
Next we compute photon statistics. Poisson statistics naturally results and
infrared divergence can be avoided even for pointlike sources. Classical fields
produced by classical sources can be obtained if one computes coherent-state
averages of Heisenberg-picture operators. It is shown that the new form of
representation automatically smears out pointlike currents. We discuss in detail
Poincar\'e covariance of the theory and the role of Bogoliubov transformations
for the issue of gauge invariance. The representation we employ is parametrized
by a number that is related to R\'enyi's $\alpha$. It is shown that the
``Shannon limit" $\alpha\to 1$ plays here a role of a correspondence principle
with the standard regularized formalism.
\end{abstract}
\maketitle

\section{Introduction}

Dynamics of observables is given in quantum mechanics by Heisenberg equations
of motion. In quantum field theory one does not really work with Heisenberg
equations but prefers the $S$-matrix formalism. The situation is caused mainly
by the fact that quantum fields are operator valued distributions and their
products taken at the same point in space-time are meaningless.
Formal solutions, when inserted back into Heisenberg equations, typically lead
to mathematical absurdities due to divergences. The difficulties are less
visible and thus easier to live with at the $S$-matrix level. Another reason
for the popularity of the $S$-matrix formalism is the question of gauge
invariance. There are various `proofs' in the literature that the $S$-matrix
formalism of quantum electrodynamics (QED) is manifestly gauge invariant.
However, as discussed in detail in \cite{IBB1967}, all the `easy
one-line proofs', including the celebrated Feynman one \cite{Feynman}, are
false if divergent expressions are encountered.

As often stressed by Dirac, the removal-of-divergences techniques practically
mean a departure from the Heisenberg equation. Dirac regarded this as a
serious drawback of quantum field theory. In his last published lecture
\cite{Dirac1984a}, entitled
`The requirements of fundamental physical theory', and  given to a gathering
of Nobel Laureates at Lindau on July 1, 1982, he said:

``I feel that we have to insist on the validity of this Heisenberg equation.
This is the whole basis of quantum theory. We have got to hold onto it
whatever we do, and if the equation gives results which are not correct
it means that we are using the wrong Hamiltonian. This is the point I
want to emphasise (...). Heisenberg originally formulated these equations
with the dynamical variables appearing as matrices. You can generalise this
very much by allowing more general kinds of quantities for your dynamical
variables. They can be any algebraic quantities such that you do not in
general have commutative multiplication (...). Some day people will find the
correct Hamiltonian and there will be some new degrees of freedom, something
which we cannot understand according to classical ideas, playing a role in
the foundations of quantum mechanics."

The talk appeared in a published form in 1984, the year of Dirac's death.
Also in 1984 Dirac published his last paper \cite{Dirac1984b}, entitled
`The future of atomic physics' where he criticised, again from the
perspective of the divergences, the dogma of irreducibility of
representations of elementary quantum symmetries.
Taken together, the two papers form a kind of scientific testament
of this great physicist.

Quite recently, in a series of papers \cite{I,II,III}, one of us
was advocating the idea that the occurrence of the divergences may
be related to the fact that the standard quantization scheme is based
on irreducible representations of canonincal (anti-)commutation
relations. It was argued that there are physical reasons for the use
of certain reducible representations of CCR (in \cite{I,II}) and
CAR (in \cite{III}).
There is no problem with multiplying fields at the same space-time
point because the new representation leads to fields that are
operators and not operator-valued distributions. Preliminary analysis
of interactions with charges discussed in these papers (two-level
atoms \cite{I}, classical current \cite{II}, reducibly quantized
fermionic fields \cite{III}) was always pointing into the conclusion
that the reducible representation {\it may\/} indeed remove
divergences, although definite statements would have been far
premature at that stage.

Particularly intriguinging seems the possibility, discussed in \cite{IV,V}, that the experiments on Rabi oscillations performed by the Paris group \cite{Haroche} may be more consistently interpreted in terms of fields quantized by means of the reducible representations. If this were the case, it would mean that the fundamental question which representation is more physical would be within the reach of present-day optical experiments. But then we have to go further with the analysis of the new representation. In particular, we have to control the relativistic and gauge properties of the formalism, a fact that justifies the origin of the present paper.

The goal of the present work is to apply a covariant analogue of the
representation introduced in \cite{II} to the problem of
Heisenberg-picture evolution of electromagnetic fields interacting
with a classical current.
The issue of infinities is not our main concern here since any
reliable discussion of ultraviolet divergences would require
quantized currents, but various automatic regularizations due to our
choice of representation can also be observed. In the present paper
we are more interested in the problem of Poincar\'e covariance
vs.~gauge invariance, positive-definiteness of the scalar product,
and unitarity of evolution. Interaction with classical currents does
not grasp all the possible subtleties of QED, but it allows for exact
solutions and thus is a natural playground for testing any new
quantization paradigm.

The results are quite promising. As opposed to standard quantization
schemes it makes in our representation perfect sense to speak of the
Heisenberg-picture evolution of field operators. We begin with
solving the equations under the assumption that  the
fields are free at timelike infinities. We make this condition precise by first taking solutions that are free at an arbitrary $t_0$, and afterwards
taking the limit $t_0\to\pm\infty$.
The results of this limiting procedure are, respectively, retarded and advanced
solutions of the Maxwell equations.
Having the solutions for all times, we can find a unitary map
$W_\pm(t,t_1)$ that maps retarded/advanced fields at time $t_1$ into
retarded/advanced fields at time $t$. Once we have this unitary map,
we can compute its generator.
We show that it differs from the usual minimal-coupling Hamiltonian.

We study in more detail the solutions corresponding to pointlike
classical sources.
Taking a relatively general form of the charge world-line we compute
the explicit form of the modified interaction Hamiltonian. The
interaction term does not contain any self-energy contribution
but we find in addition a term that describes the work performed
by the charge against free electric field.
We then show what kind of a {\it classical\/} field one arrives at
if one computes coherent-state averages of our field operators. The
resulting field looks as if it were produced by an extended current,
even though the current is pointlike. We construct the evolution
operator and consider its associated $S$-matrix. No infrared
divergence occurs and there is no problem with computing photon
statistics even for pointlike sources.

What is interesting, the different form of representation of CCR we
work with implies also a slightly modified form of the {\it free\/}
part of the Hamiltonian. The modification is subtle and is related
to some additional degree of freedom characterized by a parameter
$N$. The role of $N$ for Rabi oscillations was discussed in detail in \cite{IV} with the conclusion that finite $N$ are compatible with experiment. The ``thermodynamic limit" $N\to\infty$ plays a role of a
correspondence principle that maps the new theory into a
{\it regularized\/} form of the standard one. Effectively, the
limiting results are equivalent to the standard ones but with a
cut-off. It has to be stressed, however, that there is no cut-off
in the Hamiltonian, and the regularization is a result of the
special form of the vacuum one employs in the new representation.
We also show that the limit $N\to\infty$ is equivalent to R\'enyi's
limit $\alpha\to 1$.

The modification of the formalism we discuss seems to satisfy
standards which are not that far from the requirements proposed by
Dirac: We have new types of noncommuting dynamical variables, they
involve certain new degrees of freedom that have no counterpart in
classical electrodynamics, and their dynamics is governed by a
modified Hamiltonian. The dynamics is given by Heisenberg equations
but their solutions simultaneously satisfy Maxwell's equations with
a regularized current. The regularization is not introduced ad hoc,
but follows from the quantum structure. As opposed to standard
quantization schemes the reducibly quantized fields are less singular
than their classical counterparts. One can also think of this effect as an example of David Finkelstein's idea of `quantization as regularization' \cite{Finkelstein}.

The paper is organized as follows. We choose the Penrose-Rindler
spinor formalism since it naturally leads to null tetrads that are
implicitly present in Lorenz-gauge potentials, and play a role of
polarization vectors. We begin with clarifying links between the
nonuniqueness of Lorenz-gauge {\it classical\/} 4-potentials and
equivalence classes of spin frames associated with null 4-momenta.
The spin frames are later used in construction of two types of
momentum-dependent tetrads (associated with circular and linear
polarizations of spin-1 fields and `longitudinal' and `timelike'
polarizations of two additional scalar fields). As an intermediate
step towards quantization we explain in Section V how a change of
spin frame within an equivalence class is related to a Bogoliubov
transformation of creation and annihilation operators. For classical
fields we do not have to worry about changes within equivalence
classes since anyway the result is a gauge transformation. However,
when we quantize the 4-potential the changes of spin frames become
visible in transformations of the two `timelike' and `longitudinal'
fields and the formalism becomes ambiguous unless we compensate this
modification by a unitary transformation of annihilation and creation
operators --- this is where we need a Bogoliubov transformation.
In Section VI we introduce the reducible representation, parametrized
by a natural number $N$, of canonical commutation relations and,
in Section VII, we briefly explain why in thermodynamic limit
$N\to\infty$ the representation will automatically introduce cut-offs
at the level of averages. In Section VII we show how to quantize the
4-potential in a manifestly covariant way and without indefinite
metric or non-Hermitian field operators. In Section IX we discuss
the analogue of the Jordan-Pauli function and show that distributions
typical of the standard formalism are here replaced by non-singular
objects. Section X discusses Poincar\'e covariance of the theory.
Sections XI and XII investigate various subtleties of the Lorenz
condition. In Section XIII we introduce vacuum, multi-photon, and
coherent states. The problem of fields produced by classical currents
is discussed in Section XIV. We solve Heisenberg equations under the
assumption that at $t_0$ the fields are free, and we conclude that the
assumption makes physical sense only if $t_0$ is moved to $\pm\infty$.
In these limits the solutions of Heisenberg equations are equivalent
to retarded and advanced fields, but the effective current that occurs
in the source term is not a `$c$-number' but an operator that
nevertheless commutes with all field operators
(the operator can be nontrivial because the representation is reducible). Having these
solutions we are in position to find an explicit unitary transformation
that shifts a retarded/advanced solution in time by a finite $\Delta t$.
We find both the transformation and its generator in Section XVI, and
in Section XVII we check the result on a pointlike charge. In this way
we have systematically derived the Hamiltonian that allows to solve
the problem with retarded or advanced initial condition at $t=0$
instead of that with free field at $\pm\infty$.
In Section XVIII we address the issue of $S$-matrix and the associated
photon statistics. Finally, having the operator solutions we compute
their coherent-state averages and discuss the implications of the
formalism for classical electrodynamics.

\section{Equivalence classes of spin-frames and classical gauge freedom}

The 4-momentum $k^a=k^a(\bm k)=(|\bm k|,\bm k)$ of a massless particle
can be written in spinor notation \cite{PR} as
$k^a=\pi^A(\bm k)\bar \pi^{A'}(\bm k)$, where
$\pi^A(\bm k)$ is a spinor field defined by $k^a$ up to a phase factor.
For any $\pi^A(\bm k)$ there exists another spinor $\omega^A(\bm k)$
satisfying the spin-frame condition $\omega_A(\bm k)\pi^A(\bm k)=1$.
Given $\pi^A(\bm k)$ we cannot find a unique $\omega^A(\bm k)$, since
for any function $\phi(\bm k)$ the new field
\be
\tilde \omega^A(\bm k)=\omega^A(\bm k)+\phi(\bm k)\pi^A(\bm k)
\label{sim}
\ee
also satisfies
$\tilde \omega_A(\bm k)\pi^A(\bm k)=1$. This leads to the equivalence relation:
$\tilde \omega^A(\bm k)\sim \omega^A(\bm k)$ iff
$\tilde \omega^A(\bm k)-\omega^A(\bm k)$ is proportional to
$\pi^A(\bm k)$.

Free classical electromagnetic fields are related to $\pi^A(\bm k)$ by
\be
{F}_{ab}(x)
&=&
\partial_a A_b(x)-\partial_b A_a(x)=
-{\textstyle\int} d\Gamma(\bm k)
\pi_A(\bm k)\pi_B(\bm k)\varepsilon_{A'B'}
\big(\alpha(\bm k,-)e^{-ik\cdot x}
+\overline{\alpha(\bm k,+)}e^{ik\cdot x}\big)
+ {\rm c.c.}\label{F}
\ee
where $d\Gamma(\bm k)$ is the invariant measure on the light-cone.
The 4-potential in a Lorenz \cite{Lorenz} gauge can be taken in the form
(cf.~\cite{Ashtekar})
\be
A_a(x) &=& i{\textstyle\int} d\Gamma(\bm k)e^{-ik\cdot x}
\big(
\omega_A(\bm k)\bar\pi_{A'}(\bm k)
\alpha(\bm k,+)
+
\pi_{A}(\bm k)\bar\omega_{A'}(\bm k)
\alpha(\bm k,-)
\big)
+ {\,\rm c.c.}\label{A}
\ee
Now, if we replace $\omega^A(\bm k)$ by $\tilde \omega^A(\bm k)$
belonging to the same equivalence class, i.e. satisfying (\ref{sim}),
then
\be
{A}_a(x) &\mapsto&
\tilde {A}_a(x)
=
{A}_a(x)
-
\partial_a
\Phi(x)
\ee
where
\be
\Phi(x)
&=&
{\textstyle\int} d\Gamma(\bm k)
\phi(\bm k)
\big(\alpha(\bm k,+)e^{-ik\cdot x}
+
\overline{\alpha(\bm k,-)}e^{ik\cdot x}
\big)
+\,{\rm c.c.}
\ee
is a solution of $\Box\Phi(x)=0$. It follows that the equivalence
class of spin-frames corresponds to an equivalence class of
Lorenz-gauge potentials.

\section{Tetrads associated with spin-frames}

Following \cite{PR} we introduce two types of tetrads associated
with the spin-frames.
The null tetrad employs the 4-momentum as one of its elements:
\be
k_{a}(\bm k)
&=&
\pi_{A}(\bm k)\bar\pi_{A'}(\bm k)\\
\omega_{a}(\bm k)
&=&
\omega_{A}(\bm k)\bar\omega_{A'}(\bm k)\\
m_{a}(\bm k)
&=&
\omega_{A}(\bm k)\bar\pi_{A'}(\bm k)\\
\bar m_{a}(\bm k)
&=&
\pi_{A}(\bm k)\bar\omega_{A'}(\bm k).
\ee
Complex 4-vectors $m_{a}(\bm k)$, $\bar m_{a}(\bm k)$ occur in
(\ref{A}) and play the role of circular polarization vectors.
The Lorenz condition satisfied by (\ref{A}) follows from the
transversality property $k^{a}(\bm k) m_{a}(\bm k)
=k^{a}(\bm k)\bar m_{a}(\bm k)=0$. The null tetrad is related
to the Minkowski-space metric tensor of signature $(+,-,-,-)$
in the standard way \cite{PR}
\be
g_{ab}
&=&
k_a(\bm k)\omega_b(\bm k)+\omega_a(\bm k) k_b(\bm k)
-m_a(\bm k)\bar m_b(\bm k)-m_b(\bm k)\bar m_a(\bm k).\label{g-id}
\ee
This formula is independent of the choice of the representative
$\omega_{A}(\bm k)$ of a given equivalence class.

The tetrad defined by
\be
x_{a}(\bm k)
&=&
{\textstyle\frac{1}{\sqrt{2}}}
\big(m_{a}(\bm k)+\bar m_{a}(\bm k)\big)
\\
y_{a}(\bm k)
&=&
{\textstyle\frac{i}{\sqrt{2}}}
\big(m_{a}(\bm k)-\bar m_{a}(\bm k)\big)
\\
z_{a}(\bm k)
&=&
{\textstyle\frac{1}{\sqrt{2}}}
\big(\omega_{a}(\bm k)-k_{a}(\bm k)\big)
\\
t_{a}(\bm k)
&=&
{\textstyle\frac{1}{\sqrt{2}}}
\big(\omega_{a}(\bm k)+k_{a}(\bm k)\big)
\ee
is, in the terminology of \cite{PR}, a restricted Minkowski tetrad, satisfying
\be
g_{ab}
&=&
t_{a}(\bm k)
t_{b}(\bm k)
-
x_{a}(\bm k)
x_{b}(\bm k)
-
y_{a}(\bm k)
y_{b}(\bm k)
-
z_{a}(\bm k)
z_{b}(\bm k)
\ee
The potential now can be written as
\be
{A}_a(x)
&=&
i{\textstyle\int} d\Gamma(\bm k)
\big(
x_{a}(\bm k)
\alpha(\bm k,1)
 +
y_{a}(\bm k)
\alpha(\bm k,2)
\big)e^{-ik\cdot x}
+ {\rm c.c.}
\ee
The link between the two types of amplitudes
\be
\alpha(\bm k,\pm)
&=&
\textstyle{\frac{1}{\sqrt{2}}}
\big(
\alpha(\bm k,1)\pm i\, \alpha(\bm k,2)
\big)
\ee
is analogous to this between circular and linear polarizations.
The link is not accidental.

\section{Transformation properties of spin-frames}

The transformation properties of spin-frames we shall discuss
below do not depend on their explicit realization.
Let us denote by $\bm{\Lambda k}$ the spacelike part of the
4-vector
$\Lambda{_a}{^b}k_b(\bm k)$, and by $\Lambda{_A}{^B}$ the
unprimed SL(2,C) matrix corresponding to
$\Lambda{_a}{^b}\in$ SO(1,3).
The spinor-field transformation
\be
\pi_{A}(\bm k)
&\mapsto&
\Lambda\pi_{A}(\bm k)
=
\Lambda{_A}{^B}\pi_{B}(\bm{\Lambda^{-1} k})
\ee
implies that
$k_{a}(\bm k)
=
\Lambda\pi_{A}(\bm k)\overline{\Lambda\pi}_{A'}(\bm k)
$
and, hence,
\be
\Lambda\pi_{A}(\bm k)
=
e^{-i\Theta(\Lambda,k)}
\pi_{A}(\bm k)
\ee
The phase factor
\be
e^{-i\Theta(\Lambda,k)}
=
\Lambda^{AB}\omega_{A}(\bm k)\pi_{B}(\bm{\Lambda^{-1} k})
\ee
is the one occuring in the unitary spin-1/2 zero-mass
representation of the (covering space of the) Poincar\'e
group, and does not depend on the choice of the representative
$\omega_{A}(\bm k)$. The angle $\Theta(\Lambda,k)$ is
known as the Wigner phase.

An analogously defined
\be
\Lambda\omega_{A}(\bm k)
=
\Lambda{_A}{^B}\omega_{B}(\bm{\Lambda^{-1} k})
\ee
satisfies
\be
\Lambda\omega_{A}(\bm k)\Lambda\pi^{A}(\bm k)=1
\ee
and thus
\be
\Lambda\omega_{A}(\bm k)
&=&
e^{i\Theta(\Lambda,k)}\big(\omega_{A}(\bm k)
+\phi(\bm k)\pi_{A}(\bm k)\big)\\
&=&
e^{i\Theta(\Lambda,k)}\tilde\omega_{A}(\bm k)\label{Lambda tilde omega}
\ee
with some $\phi(\bm k)$. The new $\tilde\omega_{A}(\bm k)$
belongs to the equivalence class of $\omega_{A}(\bm k)$.
The gauge transformation $\omega_{A}(\bm k)
\mapsto \tilde\omega_{A}(\bm k)$ is in general nontrivial
and depends on the explicit form of the spin-frame one
works with. This is the reason why 4-potentials are not
4-vector fields: A Lorentz transformation changes Lorenz
gauges within the equivalence class of spin-frames.
In order to guarantee independence of physical quantities
of particular representatives one employs the result
described in the next Section.

\section{Equivalence classes of spin frames and
Bogoliubov transformation}
\label{Sec.Bog}

Of crucial importance for the formalism we shall develop
below is the link between the transformation (\ref{sim})
and a transformation of a Bogoliubov type, i.e.~the one
mixing creation and annihilation operators in a way that
preserves canonical commutation relations.

To begin with, consider four annihilation operators $a_j$
satisfying $[a_j,a_{j'}^{\dag}]
=\delta_{jj'}1$ and the operator
\be
V_a &=& x_a a_1+y_a a_2+z_a a_3+t_a a_0^{\dag}.
\ee
The change of spin-frame by
$\omega_A\mapsto \omega_A+|\phi|e^{-i\theta}\pi_A$
is represented at the tetrad level by
\begin{widetext}
\be
\left(
\begin{array}{c}
t_a\\
x_a\\
y_a\\
z_a
\end{array}
\right)
&\mapsto&
\left(
\begin{array}{c}
\tilde t_a\\
\tilde x_a\\
\tilde y_a\\
\tilde z_a
\end{array}
\right)
=
\left(
\begin{array}{cccc}
1+|\phi|^2/2 & |\phi|\cos\theta & |\phi|\sin\theta & -|\phi|^2/2 \\
|\phi|\cos\theta & 1 & 0 & -|\phi|\cos\theta \\
|\phi|\sin\theta & 0 & 1 & -|\phi|\sin\theta \\
|\phi|^2/2 & |\phi|\cos\theta & |\phi|\sin\theta & 1-|\phi|^2/2
\end{array}
\right)
\left(
\begin{array}{c}
t_a\\
x_a\\
y_a\\
z_a
\end{array}
\right)\label{a->b}
\ee
\end{widetext}
The matrix in (\ref{a->b}) is a special Lorentz transformation
in Minkowski space with metric tensor diag$(1,-1,-1,-1)$.
Let us now introduce four new operators $b_j$ defined implicitly by
\be
V_a &=& t_a a_0^{\dag}+x_a a_1+y_a a_2+z_a a_3= \tilde t_a b_0^{\dag}+\tilde x_a b_1+\tilde y_a b_2
+\tilde z_a b_3\label{tilde V}
\ee
The explicit form
\be
b_1
&=&
a_1 - |\phi|\cos\theta a_3 -|\phi|\cos\theta a_0^{\dag}
\nonumber\\
b_2
&=&
a_2 - |\phi|\sin\theta a_3  -|\phi|\sin\theta a_0^{\dag}\nonumber\\
b_3
&=&
|\phi|\cos\theta a_1 +|\phi|\sin\theta a_2
+ (1-{\textstyle\frac{1}{2}}|\phi|^2)a_3 -
{\textstyle\frac{1}{2}}|\phi|^2a_0^{\dag} \nonumber\\
b_0
&=&
-|\phi|\cos\theta a_1^{\dag} -|\phi|\sin\theta a_2^{\dag} +
{\textstyle\frac{1}{2}}|\phi|^2a_3^{\dag}
+ (1+{\textstyle\frac{1}{2}}|\phi|^2)a_0\nonumber
\ee
implies that $[b_j,b_{j'}^{\dag}]=\delta_{jj'}1$ and, hence,
there exists \cite{Bratelli} a unitary Bogoliubov-type
transformation $B$ satisfying $b_j=B^{\dag}a_jB$.
In order to explicitly construct $B$ \cite{Klaudia} we introduce the representation of the Lie algebra so$(1,3)$,
\be
J_1 &=& i(a_3^{\dag}a_2-a_2^{\dag}a_3),\\
J_2 &=& i(a_1^{\dag}a_3-a_3^{\dag}a_1),\\
J_3 &=& i(a_2^{\dag}a_1-a_1^{\dag}a_2),\\
K_1 &=& i(a_0^{\dag}a_1^{\dag}-a_0a_1),\\
K_2 &=& i(a_0^{\dag}a_2^{\dag}-a_0a_2),\\
K_3 &=& i(a_0^{\dag}a_3^{\dag}-a_0a_3),
\ee
and their combinations that form a representation of e$(2)$,
\be
L_1 &=& K_1+J_2=i(a_0^{\dag}a_1^{\dag}-a_0a_1+a_1^{\dag}a_3-a_3^{\dag}a_1),\\
L_2 &=& K_2-J_1=i(a_0^{\dag}a_2^{\dag}-a_0a_2-a_3^{\dag}a_2+a_2^{\dag}a_3),\\
L_3 &=& J_3,\\
{[L_1,L_3]}
&=&
-iL_2,\\
{[L_2,L_3]}
&=&
iL_1,\\
{[L_1,L_2]}
&=&
0.
\ee
Then
\be
B
&=&
B(\phi)=
e^{i|\phi|(L_1\cos\theta +L_2\sin \theta )},\label{B(phi)}
\ee
as can be verified by a straightforward computation. Obviously, $[B(\phi_1),B(\phi_2)]=0$ for any $\phi_1$, $\phi_2$, since  the map $\phi\mapsto B(\phi)$ is a representation of translations.

\section{Construction of the reducible representation of CCR}

The construction is analogous to the one introduced in \cite{II}.
The modification with respect to \cite{II} is that here we
introduce four types of annihilation operators, and not just
two corresponding to the polarization degrees of freedom.

One begins with four operators, $a_0$, $a_2$, $a_2$, $a_3$,
satisfying commutation relations typical of an
{\it irreducible\/} representation of CCR:
$[a_j,a_{j'}^{\dag}]=\delta_{jj'}1$.
Let $|0\rangle$ denote their common vacuum,
i.e.~$a_j|0\rangle=0$.
Now take the kets $|\bm k\rangle$ normalized with respect
to the light-cone delta function
\be
\langle\bm k|\bm k'\rangle
&=&
\delta_\Gamma(\bm k,\bm k')
=
(2\pi)^3 2|\bm k|\delta^{(3)}(\bm k-\bm k').
\ee
What we call the $N=1$ (or 1-oscillator) representation
of CCR acts in the Hilbert space
$\cal H$ spanned by kets of the form
\be
|\bm k,n_0,n_1,n_2,n_3\rangle
=
|\bm k\rangle\otimes
\frac{(a_0^{\dag})^{n_0}(a_1^{\dag})^{n_1}(a_2^{\dag})^{n_2}(a_3^{\dag})^{n_3}}
{\sqrt{n_0!n_1!n_2!n_3!}}|0\rangle.\nonumber
\ee
Physically, $\cal H$ may be regarded as representing the space
of states of a single four-dimensional oscillator.
The 1-oscillator representation is defined by
\be
a(\bm k,j)=|\bm k\rangle\langle\bm k|\otimes a_j.
\ee
This representation is reducible since the commutator
\be
[a(\bm k,j),a(\bm k',j')^{\dag}]
&=&
\delta_{jj'}\delta_\Gamma(\bm k,\bm k')
|\bm k\rangle\langle\bm k|\otimes 1
\ee
involves at the right-hand-side the operator-valued distribution
$I(\bm k)=|\bm k\rangle\langle\bm k|\otimes 1$ belonging to
the center of the algebra,
$[a(\bm k,j),I(\bm k')]=[I(\bm k),a(\bm k',j')^{\dag}]=0$,
for all $\bm k$, $\bm k'$, $j$, $j'$.
The
$I(\bm k)$ form a resolution of unity
\be
{\textstyle\int} d\Gamma(\bm k)I(\bm k)
&=&
{\textstyle\int} d\Gamma(\bm k)|\bm k\rangle\langle\bm k|\otimes 1=I.
\ee
Here $I$ is the identity operator in $\cal H$.

For arbitrary $N$ the representation is constructed as follows.
Let $\cal H$ be the representation space of the $N=1$
representation. Define
\be
\uu {\cal H}=\underbrace{\cal H\otimes\dots\otimes \cal H}_N
\ee
and let $A$ be an arbitrary operator defined for $N=1$. Let
\be
A^{(n)}=\underbrace{I\otimes\dots\otimes I}_{n-1}\otimes A
\otimes \underbrace{I\otimes\dots\otimes I}_{N-n}.
\ee
The $N$ oscillator extension of $a(\bm k,j)$ is defined by
\be
\uu a(\bm k,j)
&=&
\textstyle{\textstyle{\frac{1}{\sqrt{N}}}}\textstyle{\sum}_{n=1}^N a(\bm k,j)^{(n)}
\ee
and satisfies the reducible representation
\be
[\uu a(\bm k,j),\uu a(\bm k',j')^{\dag}]
&=&
\delta_{jj'}\delta_\Gamma(\bm k,\bm k')
\uu I(\bm k)
\ee
where
\be
\uu I(\bm k)
&=&
\textstyle{\frac{1}{N}}\textstyle{\sum}_{n=1}^N I(\bm k)^{(n)}.\label{uu I(k)}
\ee
As before we find the resolution of unity
\be
{\textstyle\int} d\Gamma(\bm k)\uu I(\bm k)
&=&
\uu I
\ee
where $\uu I$ is the identity operator in $\uu{\cal H}$.

\section{Thermodynamic limit and quantum law of large numbers}

The asymptotic properties of the formalism for $N\to \infty$
can be anticipated already at this stage if one recognizes in
the formula (\ref{uu I(k)}) the {\it frequency operator\/}
employed in the analysis of quantum laws of large numbers
\cite{LLN1,LLN2,LLN3,LLN4}.
Indeed, let $P_\theta$ be a projector. The frequency operator
corresponding to the random variable (proposition) represented
by $P_\theta$ reads
\be
P_{\theta,N}
&=&
\textstyle{\frac{1}{N}}\textstyle{\sum}_{n=1}^N P_\theta^{(n)},
\ee
Let $P_\theta|\theta\rangle=|\theta\rangle$ and
$|\psi\rangle=\sum_\theta \psi_\theta|\theta\rangle$ be a state.
Let $|\uu \psi\rangle=|\psi\rangle\otimes\dots\otimes|\psi\rangle$ ($N$ times).
Then the following weak law of large numbers holds true
\be
\lim_{N\to\infty}
\parallel
(P_{\theta,N})^m|\uu \psi\rangle-|\psi_\theta|^{2m}|\uu \psi\rangle
\parallel
=0
\ee
for $m=1,2,3,\dots$. The weak law states that effectively,
for $N\to\infty$, the frequency operators act by multiplication
of $N$-copy states by appropriate probabilities.
This is essentially why also in our field theory we will find
averages where, in the limit
$N\to\infty$, the operators $\uu I(\bm k)$ will be replaced by
their corresponding probabilities $Z(\bm k)$ associated with
the choice of the vacuum state.

\section{Quantization of the potential}

The potential operator at the level of the $N$-oscillator representation reads
\begin{widetext}
\be
{\uu A}_a(x)
&=&
i{\textstyle\int} d\Gamma(\bm k)
\big(
x_{a}(\bm k)
\uu a(\bm k,1)
 +
y_{a}(\bm k)
\uu a(\bm k,2)
+
z_{a}(\bm k)
\uu a(\bm k,3)
 +
t_{a}(\bm k)
\uu a(\bm k,0)^{\dag}
\big)e^{-ik\cdot x}
+ {\,\rm h.c.}\label{Axy}\\
&=&
i{\textstyle\int} d\Gamma(\bm k)
\big(
m_{a}(\bm k)
\uu a(\bm k,+)
+
\bar m_{a}(\bm k)
\uu a(\bm k,-)
+
z_{a}(\bm k)
\uu a(\bm k,3)
 +
t_{a}(\bm k)
\uu a(\bm k,0)^{\dag}
\big)e^{-ik\cdot x}
+ {\,\rm h.c.}\label{A+-}
\ee
\end{widetext}
It is better to think of (\ref{Axy}) and (\ref{A+-}) as operators
representing a system quantized at the $N=1$ level, and then
extended to arbitrary $N$ by
\be
{\uu A}_a(x)
&=&
\textstyle{\textstyle{\frac{1}{\sqrt{N}}}}
\textstyle{\sum}_{n=1}^N {A}_a(x)^{(n)}.
\ee
From such a perspective it is easier to understand the structure
of generators of the Poincar\'e group and other observables.

Let us note that the potential is Hermitian and the Hilbert space
$\uu{\cal H}$ involves a positive-definite scalar product.
A change (\ref{sim}) of spin-frame  can be compensated by an
$N$-oscillator Bogoliubov transformation
$\uu B=B\otimes\dots\otimes B$, where $B$ is of the type discussed
in Section \ref{Sec.Bog}.

The commutator of fields taken at arbitrary space-time points
\be
[{\uu A}_a(x),{\uu A}_b(y)]
&=&
ig_{ab} \uu D(x-y)\label{[A,A]}
\ee
involves the operator analogue of the Jordan-Pauli function
\be
\uu D(x)
&=&
i{\textstyle\int} d\Gamma(\bm k)
\uu I(\bm k)
\big(
e^{-ik\cdot x}
-
e^{ik\cdot x}
\big).
\label{J-P}
\ee
The correct signature of the metric tensor in (\ref{[A,A]})
comes from the Bogoliubov-type structure of the positive-frequency
part of $\uu A_a(x)$, i.e. the combination of annihilation and
creation operators. If one had replaced $a_0^{\dag}$ by $a_0$
one would have been forced to depart either from positivity of
the scalar product or unitarity of evolution. A reducible
version of such a (Gupta-Bleuler) formalism is possible \cite{MC-GB},
but one can show that contradictions with probability interpretation
of the theory would necessary occur (for a brief discussion of the
problem cf.~ Section \ref{Qway}).

\section{Jordan-Pauli operator: the roots of regularization}

To understand why the formalism we construct is less singular than
the one based on irreducible representations it is instructive to
take a closer look at (\ref{J-P}).
In the first place, formula (\ref{J-P}) is typical of all the
representations of CCR, reducible or irreducible, the differences
boiling down to different explicit forms of the central element
$\uu I(\bm k)$. The standard Pauli-Jordan function corresponds to
representations where $\uu I(\bm k)$ equals the identity. In our
representation we can write
(cf. Eq.~(\ref{uu I(k)}))
\be
\uu D(x)
&=&
\uu D^{(+)}(x)
+
\uu D^{(-)}(x)
\\
\uu D^{(\pm)}(x)
&=&
\pm i{\textstyle\int} d\Gamma(\bm k) \uu I(\bm k)e^{\mp ik\cdot x}
=
\textstyle{\frac{1}{N}}\textstyle{\sum}_{n=1}^N D^{(\pm)}(x)^{(n)}.
\label{uu D+}
\ee
The operator whose $N$-oscillator extensions occur in (\ref{uu D+})
reads explicitly
\be
D^{(\pm)}(x)
&=&
\pm i{\textstyle\int} d\Gamma(\bm k)
|\bm k\rangle\langle\bm k| e^{\mp ik\cdot x}\otimes 1
=
\pm i e^{\mp i\hat k\cdot x}\otimes 1
\hat k_a
=
{\textstyle\int}d\Gamma(\bm k) k_a|\bm k\rangle\langle\bm k|.
\ee
As we can see, the operators $D^{(\pm)}(x)$ are unitary
representations of 4-translations, and their generators are
given by $\hat k_a$. In particular,
\be
D^{(\pm)}(0)
=
\pm i I,\quad
\uu D^{(\pm)}(0)
=
\pm i \uu I.
\ee
Quantization in terms of our reducible representation replaces
distributions by unitary operators. This is the main difference
with respect to the schemes based on regularizations of
distributions \cite{NL1,NL2,NL3,NL4,NL5,NL6,NL7}. In our approach
there are no cut-off functions in Heisenberg-picture operators.
They appear effectively at the level of averages and are due to
the properties of {\it states\/} (e.g. compare the operator
(\ref{Aret}), involving the frequency operator $\uu I(\bm k)$
and no cut-off, with the average
(\ref{eff A}), involving the probability $Z(\bm k)$ and the
cut-off function $\chi(\bm k)$). As one of the consequences,
spectra of Hamiltonians occurring in reducibly quantized theories
will not depend on cut-offs.

These facts show that the regularization occuring in our approach
is of an entirely different origin than the ones we know from
quantizations based on irreducible representations.

\section{Poincar\'e covariance of free fields}
\label{Poincare}

The Poincar\'e transformations will be taken in the form
\be
\uu a(\bm {k},\pm)
&\mapsto&
e^{\pm 2i\Theta(\Lambda,\bm k)}e^{ik\cdot y}
\uu a(\bm {\Lambda^{-1}k},\pm)
= {\uu U}_{\Lambda,y}^{\dag} \uu a(\bm {k},\pm){\uu U}_{\Lambda,y}
\label{P-a}\\
\uu a(\bm {k},3)
&\mapsto&
e^{ik\cdot y}
\uu a(\bm {\Lambda^{-1}k},3)
= {\uu U}_{\Lambda,y}^{\dag} \uu a(\bm {k},3){\uu U}_{\Lambda,y}
\label{P-a3}
\\
\uu a(\bm {k},0)^{\dag}
&\mapsto&
e^{ik\cdot y}
\uu a(\bm {\Lambda^{-1}k},0)^{\dag}
= {\uu U}_{\Lambda,y}^{\dag} \uu a(\bm {k},0)^{\dag}{\uu U}_{\Lambda,y}
\label{P-a0}
\ee
where $\Theta(\Lambda,\bm k)$ is the Wigner phase. Transformation
(\ref{P-a}) is the unitary spin-1 massless representation of the
Poincar\'e group. Transformations (\ref{P-a3}), (\ref{P-a0})
imply that the additional two fields are spin-0 and massless.
Similarly to \cite{II} we reduce the construction to the
problem of finding $U_{\Lambda,y}$ satisfying
\be
{\uu U}_{\Lambda,y} &=&
\underbrace{U_{\Lambda,y}\otimes\dots\otimes U_{\Lambda,y}}_N.
\ee

\subsection{Four-translations}

The 4-momentum for $N=1$ reads
\be
P_a
&=&
{\textstyle\int} d\Gamma(\bm k)k_a|\bm k\rangle\langle \bm k|\otimes
\big(a^{\dag}_1a_1+a^{\dag}_2a_2+a^{\dag}_3a_3-a^{\dag}_0a_0\big)
\nonumber
\\
&=&
\underbrace{
{\textstyle\int} d\Gamma(\bm k)k_a|\bm k\rangle\langle \bm k|\otimes
\big(a^{\dag}_1a_1+a^{\dag}_2a_2\big)}_{P_a^I}
+
\underbrace{
{\textstyle\int} d\Gamma(\bm k)k_a J(\bm k)}_{P_a^{II}},\label{P}
\ee
with
\be
J(\bm k)=|\bm k\rangle\langle \bm k|\otimes (a^{\dag}_3a_3-a^{\dag}_0a_0)
\ee
One immediately verifies that
\be
e^{iP\cdot x}a(\bm k,\pm)e^{-iP\cdot x} &=& a(\bm k,\pm) e^{-ix\cdot k}\\
e^{iP\cdot x}a(\bm k,3)e^{-iP\cdot x} &=& a(\bm k,3) e^{-ix\cdot k}\\
e^{iP\cdot x}a(\bm k,0)^{\dag}e^{-iP\cdot x} &=& a(\bm k,0)^{\dag} e^{-ix\cdot k}
\ee
implying
\be
\uu{U}_{\bm 1,y}^{\dag}\uu A_{a}(x)\uu{U}_{\bm 1,y} &=&
\uu{A}_{a}(x-y).
\ee
The operator $J(\bm k)$
will later reappear as the generator of rotations in the group E(2)
associated with the 4-potential. The part $P_a^I$ is identical to
the 4-momentum operator introduced in \cite{II}. The 4-momentum
for arbitrary $N$ reads
\be
\uu P_a &=&\textstyle{\sum}_{n=1}^N P_a^{(n)}\label{uu P}\\
&=& \uu P_a^I+{\textstyle\int} d\Gamma(\bm k)k_a \uu J(\bm k)\label{uu P'}
\ee
Only for $N=1$ the expression coincides with the generator found by
standard Noether formulas (cf. the discussion of this point in \cite{I}).
The form (\ref{uu P}) is characteristic of a 4-momentum of $N$
non-interacting particles. These particles (four-dimensional oscillators)
have no counterpart in classical electrodynamics. It should be stressed
that these are not the oscillators of Heisenberg, Born, and Jordan
\cite{HBJ1925} since there is no relationship between $N$, which is
finite, and the number of different frequencies, which is infinite.

\subsection{Rotations and boosts}

To find an analogous representation of
\be
\uu a(\bm {k},\pm)
&\mapsto&
e^{\pm 2i\Theta(\Lambda,\bm k)}
\uu a(\bm {\Lambda^{-1}k},\pm)
= {\uu U}_{\Lambda,0}^{\dag} \uu a(\bm {k},\pm){\uu U}_{\Lambda,0}
\\
\uu a(\bm {k},3)
&\mapsto&
\uu a(\bm {\Lambda^{-1}k},3)
= {\uu U}_{\Lambda,0}^{\dag} \uu a(\bm {k},3){\uu U}_{\Lambda,0}
\\
\uu a(\bm {k},0)^{\dag}
&\mapsto&
\uu a(\bm {\Lambda^{-1}k},0)^{\dag}
= {\uu U}_{\Lambda,0}^{\dag} \uu a(\bm {k},0)^{\dag}{\uu U}_{\Lambda,0}
\ee
we take the same definition as in \cite{II}, i.e.
\be
U_{\Lambda,0} &=&
\exp\big({\textstyle\sum}_{s=\pm} 2is{\textstyle\int} d\Gamma(\bm k)
\Theta(\Lambda,\bm k)|\bm
k\rangle \langle \bm k|\otimes a^{\dag}_sa_s\big)
\big({\textstyle\int} d\Gamma(\bm p)|\bm p\rangle
\langle \bm {\Lambda^{-1}p}|\otimes  1\big).
\ee
Taking into account the properties of spin-frames and tetrads
one verifies that
\begin{widetext}
\be
\uu {U}_{\Lambda,0}^{\dag}{\uu A}_{a}(x)\uu{U}_{\Lambda,0}
&=&
i {\textstyle\int} d\Gamma(\bm k)e^{-ik\cdot \Lambda^{-1}x}
\nonumber\\
&\times&
\big(
\tilde m_{a}(\bm {\Lambda k})
e^{2i\Theta(\Lambda,\bm {\Lambda k})}\uu a(\bm {k},+)
+
\tilde {\bar m}_a(\bm {\Lambda k})
e^{-2i\Theta(\Lambda,\bm {\Lambda k})}\uu a(\bm {k},-)
+
\tilde z_{a}(\bm {\Lambda k})\uu a(\bm {k},3)
+
\tilde t_a(\bm {\Lambda k})\uu a(\bm {k},0)^{\dag}
\big)
+{\rm h.c.}
\nonumber\\
&=&
\Lambda{_a}{^b}\tilde{\uu A}_{b}(\Lambda^{-1}x)
=
\Lambda{_a}{^b}\uu B(\Lambda)^{\dag}{\uu A}_{b}(\Lambda^{-1}x)\uu B(\Lambda)
\ee
\end{widetext}
where $\uu B(\Lambda)$ compensates the change of gauge
caused by (\ref{Lambda tilde omega}).
One can costruct $\uu B(\Lambda)$ by first finding an appropriate $B(\Lambda)$ of the form analogous to (\ref{B(phi)}), and then defining $\uu B(\Lambda)=B(\Lambda)^{\otimes N}$.
It is more elegant to assume that  $\uu{U}_{\Lambda,0}$
is accompanied by a redefinition of vacuum (see below)
$|\uu O\rangle\mapsto\uu B(\Lambda)^{\dag}|\uu O\rangle$.
Then the transformation of the potential becomes effectively
\be
\uu {U}_{\Lambda,y}^{\dag}{\uu A}_{a}(x)\uu{U}_{\Lambda,y}
=
\Lambda{_a}{^b}{\uu A}_{b}\big(\Lambda^{-1}(x-y)\big)
\ee
i.e. that of a 4-vector field.

A still simpler way is to assume the transformation rule
\be
\Lambda{_A}{^B}\omega_{B}(\bm{\Lambda^{-1}k})
&=&
e^{i\Theta(\Lambda,k)}\omega_{A}(\bm k)\label{[Lambda omega]}
\ee
i.e. to replace (\ref{Lambda tilde omega}) by the form typical
of the entire equivalence class. Then one can put $\uu B(\Lambda)=\uu I$.

The operators ocuring at right-hand-sides of field commutators
transform as translation invariant scalar fields
\be
\uu {U}_{\Lambda,y}^{\dag}\uu I(\bm k)\uu{U}_{\Lambda,y}
&=&
\uu I(\bm{\Lambda^{-1}k})\\
\uu {U}_{\Lambda,y}^{\dag}\uu D(x)\uu{U}_{\Lambda,y}
&=&
\uu D(\Lambda^{-1}x).
\ee
The representation we have introduced is a direct sum of a massless, spin-1 unitary representation (corresponding to the indices 1 and 2) and a massless spin-0 unitary representation (corresponding to the indices 3 and 0) of the Poincar\'e group. In such a structure the `longitudinal' and `timelike' components do not transform as parts of a four-vector, but as two scalar fields. However, there is a lot of freedom here. There exists, for example, an interesting representation which employs the link between Bogoliubov and SO$(1,3)$ transformations, and where all the four components behave as those of a four-vector. A particular example of this link was employed in Sec.~V. It is interesting to compare the two representations in the context of field quantization, but this will be done in a separate paper \cite{CW}.

\section{Lorenz condition and Euclidean group}

The field tensor
$\uu F_{ab}(x)=\partial_a \uu A_b(x)-\partial_b \uu A_a(x)$
consists of two parts
corresponding to spin-1 and spin-0 fields
(formulas (\ref{spin-1}) and  (\ref{spin-0}), respectively)
\begin{widetext}
\be
\uu F_{ab}(x)
&=&
-{\textstyle\int} d\Gamma(\bm k)
\pi_A(\bm k)\pi_B(\bm k)\varepsilon_{A'B'}
\big(\uu a(\bm k,-)e^{-ik\cdot x}
+\uu a(\bm k,+)^{\dag}e^{ik\cdot x}\big)
+ {\rm h.c.}\label{spin-1}\\
&-&
\sqrt{2}
{\textstyle\int} d\Gamma(\bm k)
^*M_{ab}(\bm k)
\big(
\uu\Pi_1(\bm k)\cos{k\cdot x}
+
\uu\Pi_2(\bm k)\sin{k\cdot x}
\big)\label{spin-0}
\ee
\end{widetext}
\be
^*M_{ab}(\bm k)
&=&
-k_a\omega_{b}(\bm k)+k_b\omega_{a}(\bm k)
\label{^*M}\\
\uu\Pi_1(\bm k)
&=&
\textstyle{\frac{1}{2}}\big(
\uu a(\bm k,3)
+
\uu a(\bm k,3)^{\dag}
+
\uu a(\bm k,0)
+
\uu a(\bm k,0)^{\dag}
\big)\nonumber\\
\uu \Pi_2(\bm k)
&=&
\textstyle{\frac{1}{2i}}\big(
\uu a(\bm k,3)
-
\uu a(\bm k,3)^{\dag}
+
\uu a(\bm k,0)^{\dag}
-
\uu a(\bm k,0)
\big)\nonumber
\ee
The tensor $^*M_{ab}(\bm k)$ is the dual of
\be
M_{ab}(\bm k)
&=&
i\pi_{(A}\omega_{B)}\varepsilon_{A'B'}
-
i\bar\pi_{(A'}\bar\omega_{B')}\varepsilon_{AB}\label{M}\\
k^b{}^*M_{ab}(\bm k)
&=&
-k_a\label{PL}
\ee
One immediately recognizes in (\ref{M}) and (\ref{PL})
spinor formulas for a massless angular momentum tensor and
the Pauli-Lubanski vector of helicity $-1$ (cf. \cite{PR2}, Eq.~(6.3.2)).

The gauge transformation (\ref{sim}) influences the part
(\ref{spin-0})  in $\uu F_{ab}(x)$ according to
\be
^*M_{ab}(\bm k)
&\mapsto&
^*M_{ab}(\bm k)
-k_a(\bm k)q_b(\bm k)+k_b(\bm k)q_a(\bm k)\\
q_a(\bm k)
&=&
\phi(\bm k)\bar m_a(\bm k)
+\bar \phi(\bm k) m_a(\bm k)
\ee
that is, in a way typical of angular momentum. The 4-vector
$q_a(\bm k)$ can be used to reexpress the gauge transformed spin-frame as a twistor \cite{PR2}
\be
\tilde \pi_A(\bm k)
&=&\pi_A(\bm k)\\
\tilde \omega_A(\bm k)
&=&
\omega_A(\bm k)
+
q_{AA'}(\bm k)\bar \pi^{A'}(\bm k).
\ee
Mutual relations within an equivalence class are thus
determined by the twistor equation. Change of  origin
in the space of coordinates $q_a$ can be compensated
by the Bogoliubov transformation
$\uu B$.

Together with $\uu J(\bm k)$ occuring in (\ref{uu P'})
we obtain the algebra $e(2)$
\be
{[\uu\Pi_1(\bm k),\uu J(\bm k')]}
&=&
i\delta_\Gamma(\bm k,\bm k')
\uu \Pi_2(\bm k)\\
{[\uu J(\bm k),\uu \Pi_2(\bm k')]}
&=&
i\delta_\Gamma(\bm k,\bm k')
\uu \Pi_1(\bm k)\\
{[\uu \Pi_1(\bm k),\uu \Pi_2(\bm k')]}
&=&
0
\ee
It is interesting that the removal of the scalar fields by the constraint
\be
\langle \Psi'|\uu \Pi_1(\bm k)|\Psi\rangle
=
\langle \Psi'|\uu \Pi_2(\bm k)|\Psi\rangle
=0\label{not too strong}
\ee
is analogous to the condition leading to the classical
Maxwell field if one starts from induced representations
\cite{Ohnuki}. Indeed, all massless discrete-spin
representations are found if one requires that the two
translation generators of $e(2)$ annihilate vectors from
the representation space. It might be therefore tempting
to impose the stronger constraint
\be
\uu \Pi_1(\bm k)|\Psi\rangle
=
\uu \Pi_2(\bm k)|\Psi\rangle
=0\label{too strong}
\ee
also here. To see why this condition would be too strong we
write the potential in terms of
$e(2)$:
\begin{widetext}
\be
\uu A_a(x)
&=&
-
\sqrt{2}
{\textstyle\int} d\Gamma(\bm k)
\omega_{a}(\bm k)
\big(
\uu \Pi_2(\bm k)\cos k x-\uu \Pi_1(\bm k)\sin k x
\big)-
\sqrt{2}
{\textstyle\int} d\Gamma(\bm k)
k_{a}\big(
\uu Q_1(\bm k)\cos k x
+
\uu Q_2(\bm k)\sin k x
\big)+\dots
\ee
\end{widetext}
where
\be
\uu Q_1(\bm k)
&=&
-\textstyle{\frac{1}{2i}}\big(
\uu a(\bm k,3)
-
\uu a(\bm k,3)^{\dag}
-
\uu a(\bm k,0)^{\dag}
+
\uu a(\bm k,0)
\big)\nonumber\\
\uu Q_2(\bm k)
&=&
-\textstyle{\frac{1}{2}}\big(
\uu a(\bm k,3)
+
\uu a(\bm k,3)^{\dag}
-
\uu a(\bm k,0)^{\dag}
-
\uu a(\bm k,0)
\big)\nonumber
\ee
and the dots stand for the part involving only the spin-1 fields.
The part involving $\uu {\bm Q}(\bm k)$ is a gauge term and this
is why we do not see it in $\uu F_{ab}(x)$. The entire algebra reads
\be
{[\uu \Pi_1(\bm k),\uu J(\bm k')]}
&=&
i\delta_\Gamma(\bm k,\bm k')\uu \Pi_2(\bm k)\\
{[\uu J(\bm k),\uu \Pi_2(\bm k')]}
&=&
i\delta_\Gamma(\bm k,\bm k')\uu \Pi_1(\bm k)\\
{[\uu Q_1(\bm k),-\uu J(\bm k')]}
&=&
i\delta_\Gamma(\bm k,\bm k')\uu Q_2(\bm k)\\
{[-\uu J(\bm k),\uu Q_2(\bm k')]}
&=&
i\delta_\Gamma(\bm k,\bm k')\uu Q_1(\bm k)\\
{[\uu Q_1(\bm k),\uu \Pi_1(\bm k')]}
&=&
i\delta_\Gamma(\bm k,\bm k') \uu I(\bm k)
\\
{[\uu Q_2(\bm k),\uu \Pi_2(\bm k')]}
&=&
i\delta_\Gamma(\bm k,\bm k')  \uu I(\bm k)
\\
{[\uu \Pi_1(\bm k),\uu \Pi_2(\bm k')]}
&=&
0\\
{[\uu Q_1(\bm k),\uu Q_2(\bm k')]}
&=&
0
\\
{[\uu Q_1(\bm k),\uu \Pi_2(\bm k')]}
&=&
0
\\
{[\uu Q_2(\bm k),\uu \Pi_1(\bm k')]}
&=&
0
\ee
This is the Lie algebra of the 2-dimensional Euclidean group in a phase space.
The ``position operators" $\uu{\bm Q}(\bm k)$ shift the ``momenta"
$\uu{\bm \Pi}(\bm k)$ and the constraint (\ref{too strong}) must
be inconsistent with dynamics. There is no problem with (\ref{not too strong}).

The 4-divergence of the potential
\be
\partial^a \uu A_a(x)
&=&
\sqrt{2}
{\textstyle\int} d\Gamma(\bm k)
\big(
\uu \Pi_2(\bm k)\sin k x+\uu \Pi_1(\bm k)\cos k x
\big)\nonumber
\ee
shows that the weak Lorenz condition
\be
\langle \Psi'|\partial^a \uu A_a(x)|\Psi\rangle=0\label{Lorenz}
\ee
is equivalent to (\ref{not too strong}).

\section{Lorenz condition and Poincar\'e covariance of states}

The 1-oscillator Hilbert space $\cal H$ consists of vectors
\be
|\Psi\rangle &=&
{\textstyle\sum}_{n_0,n_1,n_2,n_3=0}^\infty
{\textstyle\int} d\Gamma(\bm k)
\Psi(\bm k,n_0,n_1,n_2,n_3)|\bm k,n_0,n_1,n_2,n_3\rangle
\ee
satisfying
\be
{\textstyle\sum}_{n_0,n_1,n_2,n_3=0}^\infty
{\textstyle\int} d\Gamma(\bm k)|\Psi(\bm k,n_0,n_1,n_2,n_3)|^2
<\infty
\ee
Subspaces consisting of vectors of the form
\be
|\Psi_{n_0,n_1,n_2,n_3}\rangle &=&
{\textstyle\int} d\Gamma(\bm k)
\Psi(\bm k,n_0,n_1,n_2,n_3)|\bm k,n_0,n_1,n_2,n_3\rangle
\nonumber
\ee
are invariant subspaces of the representation constructed in Section \ref{Poincare}.

In particular, all the vectors of the form
\be
|\Psi\rangle &=&
\textstyle{\sum}_{n_1,n_2=0}^\infty
{\textstyle\int} d\Gamma(\bm k)
\Psi(\bm k,n_0,n_1,n_2,n_3)|\bm k,n_0,n_1,n_2,n_3\rangle
\ee
belong to the Poincar\'e-invariant subspace satisfying the weak Lorenz condition
(\ref{not too strong}) for $N=1$. Moreover, if we additionally require $n_0=n_3$ then
\be
\langle\Psi'|P_a^{II}|\Psi\rangle=0
\ee
An extension to arbitrary $N$ is immediate. One concludes that the Lorenz condition
(\ref{not too strong}) can be imposed in a Poincar\'e invariant way.

\section{Vacuum, multiphoton, and coherent states}

The subspace corresponding to $n_0=n_1=n_2=n_3=0$ defines the vacuum for $N=1$.
Any vector of the form
\be
|O\rangle={\textstyle\int} d\Gamma(\bm k)O(\bm k)|\bm k,0\rangle
\ee
plays a role of a 1-oscillator vacuum. For arbitrary $N$ the vacuum state is taken in the form
\be
|\uu O\rangle
=
\underbrace{|O\rangle\otimes \dots\otimes |O\rangle}_N\label{uu O}
\ee
All vacuum states are annihilated by all annihilation operators.
Vacuum states are translation invariant and SL(2,C) covariant:
\be
U_{\Lambda,y}|O\rangle
&=&
{\textstyle\int} d\Gamma(\bm k)O(\bm{\Lambda^{-1} k})|\bm k,0\rangle\\
\uu U_{\Lambda,y}|\uu O\rangle
&=&
U_{\Lambda,y}|O\rangle
\otimes \dots\otimes
U_{\Lambda,y}|O\rangle
\ee
Of particular importance is the scalar field representing vacuum probability density
$Z(\bm k)=|O(\bm k)|^2$. Square integrability implies that
$Z(\bm k)$ decays at infinity;
later on, we will also require $Z(\bm k)$ going to zero at $\bm k=0$
in order to avoid infrared
divergences. The latter would spoil the explicit constructive nature
of the present approach.
The number $Z=\max_k\{Z(\bm k)\}$ is Poincar\'e
invariant and can be interpreted as a renormalization constant.

Multiphoton states are obtained in the usual way by acting on
the vacuum $|\uu O\rangle$ with creation operators. Coherent
states associated with amplitudes $\alpha(\bm k,\pm)$ occurring
in (\ref{F}) are defined in terms of the displacement operator
\cite{II}
\be
{\uu {\cal D}}(\alpha) &=& \exp\big(\uu a(\alpha)^{\dag}-\uu a(\alpha)\big)\\
\uu a(\alpha) &=&
\textstyle{\sum}_{s=\pm}{\textstyle\int} d\Gamma(\bm k)\overline{\alpha(\bm k,s)}\uu a(\bm k,s)
\ee
A coherent state is constructed from vacuum by
$|\uu O(\alpha)\rangle={\uu {\cal D}}(\alpha)|\uu O\rangle$.
Coherent-state averages are related to classical fields by
\begin{widetext}
\be
\langle \uu O(\alpha)|\uu A_a(x)|\uu O(\alpha)\rangle
&=&
i{\textstyle\int} d\Gamma(\bm k)
Z(\bm k)
\big(
m_a(\bm k)
\alpha(\bm k,+)
+
\bar m_a(\bm k)
\alpha(\bm k,-)
\big)
e^{-ik\cdot x}
+ {\,\rm c.c.}
\label{<A>}
\ee
\end{widetext}
Let us note that the averages involve the amplitudes $Z(\bm k)\alpha(\bm k,\pm)$ and not just $\alpha(\bm k,\pm)$.

\section{Fields produced by a classical current}

The interaction Hamiltonian in the interaction picture is assumed in the usual form
\be
H(t)&=&{\textstyle\int} d^3x\, J^a(t,\bm x) \uu A_a(t,\bm x)\label{H(t)}
\ee
where $J^a(t,\bm x) $ is a classical conserved current.
The interaction picture evolution operator satisfies
\be
i
\textstyle{\frac{d}{dt}}U(t,t_0)
= H(t)U(t,t_0),\quad
U(t_0,t_0)
=
\uu I
\ee
Recalling that $\uu A_a(t,\bm x)$ depends on time via the free Hamiltonian $H_0=\uu P_0$
(\ref{uu P}) we can split the Heisenberg-picture time evolution into parts involving separately
the interaction picture $U(t,t_0)$ and the free evolution, i.e.
\be
\uu A_a^H(x)=U(t,t_0)^{\dag}\uu A_a(x)U(t,t_0)\label{A^H}
\ee
To obtain the latter we made the usual assumption that there
exists a time $t_0$ at which the field is free. This restriction
is eased later on by moving $t_0$ to $\pm\infty$.
The following two splittings of (\ref{H(t)}) are important
\begin{widetext}
\be
H(t)
&=&
H_1(t)+H_1(t)^{\dag}
=
H_2(t)+H_2(t)^{\dag}\nonumber\\
H_1(t)
&=&
i
{\textstyle\int} d^3x\, J^a(x)
{\textstyle\int} d\Gamma(\bm k)
\big(
x_{a}(\bm k)
\uu a(\bm k,1)
 +
y_{a}(\bm k)
\uu a(\bm k,2)
+
z_{a}(\bm k)
\uu a(\bm k,3)
 +
t_{a}(\bm k)
\uu a(\bm k,0)^{\dag}
\big)e^{-ik\cdot x}
\\
H_2(t)
&=&
i
{\textstyle\int} d^3x\, J^a(x)
{\textstyle\int} d\Gamma(\bm k)
\big(
x_{a}(\bm k)
\uu a(\bm k,1)
 +
y_{a}(\bm k)
\uu a(\bm k,2)
+
z_{a}(\bm k)
\uu a(\bm k,3)
-
t_{a}(\bm k)
\uu a(\bm k,0)e^{2ik\cdot x}
\big)e^{-ik\cdot x}
\label{H_2}
\ee
\end{widetext}
since the commutators
\be
{[H_i(t_1),H_i(t_2)]}&=&0\label{CBH1}\\
{[H_i(t_1)^{\dag},H_i(t_2)^{\dag}]}&=&0\\
{\big[H_i(t_1),[H_j(t_2),H_j(t_3)^{\dag}]\big]}&=&0\\
{\big[H_i(t_1)^{\dag},[H_j(t_2),H_j(t_3)^{\dag}]}&=&0\label{CBH4}
\ee
hold for all $i,j=1,2$ and arbitrary times. The commutators
\be
{[H_1(x_0),H_1(y_0)^{\dag}]}
&=&
i{\textstyle\int} d^3xd^3y\, J_a(x)\uu D^{(+)}(x-y) J^a(y)
\label{H1H1}
\ee
\begin{widetext}
\be
{[H_2(x_0),H_2(y_0)^{\dag}]}
&=&
i{\textstyle\int} d^3xd^3y\, J_a(x)\uu D^{(+)}(x-y) J^a(y)
\nonumber\\
&\pp=&
+
2{\textstyle\int} d^3xd^3y\, J^a(x)  J^b(y)
{\textstyle\int} d\Gamma(\bm k) \uu I(\bm k)
t_{a}(\bm k)
t_{b}(\bm k)\cos k\cdot(x-y),
\label{H2H2}
\ee
are in the center of CCR. Employing continuous Baker-Hausdorff formulas \cite{BBMP,IZBB}
\be
T\exp\big({\textstyle\int}_{t_0}^t d\tau\big(A(\tau)+B(\tau)\big)\big)
&=&
\exp\big({\textstyle\int}_{t_0}^t d\tau (A(\tau)+B(\tau))\big)
\exp\big(\textstyle{\frac{1}{2}}{\textstyle\int}_{t_0}^t d\tau_1{\textstyle\int}_{t_0}^t
d\tau_2
\big(
\theta(\tau_1-\tau_2)-\theta(\tau_2-\tau_1)
\big)
[A(\tau_1),B(\tau_2)]\big)\nonumber\\
&=&
\exp\big({\textstyle\int}_{t_0}^t d\tau A(\tau)\big)
\exp\big({\textstyle\int}_{t_0}^t d\tau B(\tau)\big)
\exp\big({\textstyle\int}_{t_0}^t d\tau_1 {\textstyle\int}_{t_0}^t d\tau_2
\theta(\tau_1-\tau_2)[B(\tau_1),A(\tau_2)]\big)\label{cbh2}
\ee
where $A(\tau)$, $B(\tau)$ satisfy relations analogous to (\ref{CBH1})--(\ref{CBH4}), we find that
\be
U(t,t_0)
&=&
\exp\Big(-i{\textstyle\int}_{t_0}^t d^4x J^a(x) \uu A_a(x)\Big)
\exp
\Big(
-{\textstyle\frac{i}{2}\int}_{t_0}^t{\textstyle\int}_{t_0}^t d^4x_1 d^4x_2
J_a(x_1)
\uu D_{\rm adv}(x_1-x_2)
J^a(x_2)
\Big),\label{U(t,t_0)}
\ee
\end{widetext}
where $\uu D_{\rm adv}(x)
=
-\theta(-x_0)\uu D(x)$.
Formula
(\ref{cbh2}) will be later used to compute the photon statistics.

Employing (\ref{U(t,t_0)}) we find the explicit form of the Heisenberg-picture evolution
\be
\uu A_a^H(x)
&=&
\uu A_a(x)
+
\textstyle{{\textstyle\int}_{t_0}^t} d^4y \uu D(x-y)J_a(y)
\label{uu A^H}
\ee
Our field $\uu A_a^H(x)$ is free at $t=t_0$. In the next
Section we show that the weak Lorenz condition holds for (\ref{uu A^H}) only
in the limit $t_0=\pm\infty$.

\section{Lorenz condition and retarded/advanced solutions}

The four divergence of (\ref{uu A^H}) takes the form
\be
\partial^a \uu A_a^H(x)
&=&
{\rm free\,part}+
{\textstyle\int}d^3x'\uu D(t-t_0,\bm x-\bm x')J_0(t_0,\bm x')
\nonumber
\ee
Taking an arbitrary coherent-state average
\be
\langle \uu O(\alpha)|
\partial^a \uu A_a^H(x)
|\uu O(\alpha)\rangle=
{\textstyle\int}d^3x'
\langle \uu O|\uu D(t-t_0,\bm x-\bm x')|\uu O\rangle J_0(t_0,\bm x')
\nonumber
\ee
and requiring the Lorenz gauge for all conserved currents,
we obtain a condition on the vacuum-state probability density
$Z(\bm k)=|O(\bm k)|^2$
\be
\langle\uu O|\uu D(t-t_0,\bm  x)|\uu O\rangle
 =
i
\textstyle{{\textstyle\int}} d\Gamma(\bm k)Z(\bm  k)
\big(e^{-ik\cdot x}e^{ik_0 t_0}-e^{-ik_0 t_0}e^{ik\cdot x}\big)=0
\nonumber
\ee
This cannot hold in general if $t_0$ is finite. However,
for $t_0\to\pm\infty$ the condition becomes equivalent to
\be
\lim_{t_0\to\pm\infty}
\textstyle{{\textstyle\int}} d^3kf(\bm k)e^{i|\vec k| t_0}
=0
\ee
where $f(\bm k)=Z(\bm  k)e^{ik\cdot x}/|\bm k|$. The latter
condition requires only that  $Z(\bm  k)/|\bm k|$ satisfies
assumptions of the Riemann-Lebesgue lemma.

We thus restrict the analysis to the two cases
of either retarded or advanced solutions. The formulas are
\be
\uu A_a^{\rm ret}(x)
&=&
\uu A_a(x)
+
\textstyle{{\textstyle\int}} d^4y \uu D_{\rm ret}(x-y)J_a(y)
\label{Aret}\\
\uu A_a^{\rm adv}(x)
&=&
\uu A_a(x)
+
\textstyle{{\textstyle\int}} d^4y \uu D_{\rm adv}(x-y)J_a(y)
\label{Aadv}\\
\uu D_{\rm ret}(x)
&=&
\theta(x_0)\uu D(x)\\
\uu D_{\rm adv}(x)
&=&
-\theta(-x_0)\uu D(x)\\
\uu D(x)
&=&
\uu D_{\rm ret}(x)
-
\uu D_{\rm adv}(x)
\ee
Since $\Box \uu D(x)=0$ we find
\be
\Box \uu D_{\rm ret}(x-y)
=
\Box \uu D_{\rm adv}(x-y)
\stackrel{\rm def}{=}
\uu \delta(x-y)\label{uu delta}
\ee
One has to bear in mind that $\uu \delta(x-y)$ is
{\it defined\/} by (\ref{uu delta}) and that the
resulting {\it operator\/} is not equivalent to the Dirac delta.

The advanced and retarded potentials satisfy
\be
\Box \uu A_a^{\rm ret/adv}(x)
=
\textstyle{{\textstyle\int}} d^4y \uu \delta(x-y)J_a(y)
\stackrel{\rm def}{=}
\uu J_a(x)
\ee
The weak Lorenz condition implies that the average current
\be
\langle J_a(x)\rangle
=
\langle \uu O(\alpha)|
\uu J_a(x)
|\uu O(\alpha)\rangle
=
\langle \uu O|
\uu J_a(x)
|\uu O\rangle
\ee
is the conserved physical current that produces the classical electromagnetic field
\be
\langle A^{\rm ret/adv}_{a}(x)\rangle
&=&
\langle \uu O(\alpha)|
\uu A_a^{\rm ret/adv}(x)
|\uu O(\alpha)\rangle.
\ee
The modification of the current depends only on the choice of
the vacuum state because the displacement operator commutes with $\uu I(\bm k)$.

To close this Section let us mention that an operator analogue of the Feynman propagator
\be
\uu D_F(x)
&=&
\theta(x_0)\uu D^{(+)}(x)
-
\theta(-x_0)\uu D^{(-)}(x)
\\
&=&
\uu D_{\rm adv}(x)
+
\uu D^{(+)}(x)
=
\uu D_{\rm ret}(x)
-
\uu D^{(-)}(x)
\ee
would occur in {\it perturbative\/} formulas in exactly the same places as in the standard formalism. The reason is that the algebraic structure of Feynman diagrams is unchanged by the change of representation of CCR. Since $\Box\uu D^{(\pm)}(x)=0$, the Feynman potential
\be
\uu A^F_a(x)
&=&
\uu A_a(x)
+
{\textstyle\int} d^4 y\uu D_F(x-y)J_a(y)
\ee
satisfies the same equation as the retarded and advanved fields, but
is non-Hermitian for real currents.

\section{Dynamics of retarded and advanced solutions between two finite times}

We have solved the Heisenberg equations with free-field
``initial" conditions at
$t_0=\pm\infty$ and arrived at retarded and advanced
solutions of Maxwell's equations. We have not yet shown
what kind of dynamics will map retarded or advanced
solutions at a {\it finite\/} time $t_1$ into retarded
or advanced solutions at another finite time $t$. This
would be the true solution of the Heisenberg-picture
evolution since at a finite initial time the field
cannot be free, unless the charge of the current is zero.

One can immediately write down appropriate formulas on
the basis of the retarded and advanced solutions
\begin{widetext}
\be
\uu A_a^{\rm ret/adv}(t,\bm x)
&=&
\lim_{t_0\to -\infty/+\infty}
U(t,t_0)^{\dag}U_0(t,t_1)^{\dag}U(t_1,t_0)
\uu A_a^{\rm ret/adv}(t_1,\bm x)
U(t_1,t_0)^{\dag}
U_0(t,t_1)U(t,t_0)\label{W-1}
\\
&=&
W_{-/+}(t,t_1)^{\dag}
\uu A_a^{\rm ret/adv}(t_1,\bm x)W_{-/+}(t,t_1)\label{W-3}
\ee
\end{widetext}
Let us recall that $U_0(t,t_1)=\exp\big(-iH_0(t-t_1)\big)$,
where $H_0=\uu P_0$ is the free Hamiltonian defined by the
reducible representation of CCR, and $U(t,t_0)$ is the
interaction-picture evolution operator.
Some care is needed in the definitions of $W_\pm(t,t_1)$ if the limits
$
\lim_{t_0\to\pm\infty}U(t_1,t_0)^{\dag}U_0(t,t_1)U(t,t_0)
$
involve divergent phase factors. This is the standard
problem and has nothing to do with the divergences of
quantum field theory. Keeping this subtlety in mind we arrive at
\begin{widetext}
\be
W_\pm(t,t_1)
&=&
\exp\Big(i{\textstyle\int}_{\pm\infty}^{t_1} d^4x
\big(J^a(x_0,\bm x)-J^a(x_0+t-t_1,\bm x)\big)
\uu A_a(x_0,\bm x)\Big)
\exp\Big(-i H_0(t-t_1)\Big)\label{W}
\ee
\end{widetext}
It is clear that for a static charge density the evolution is free.
The ranges of integration are finite also in case the
currents are static for $t<t_-$ and
$t>t_+$ with some $t_\pm$. It should be stressed that
this type of ``switching on and off" of the current is
perfectly consistent with charge conservation.

The corresponding Hamiltonian $H_\pm(t)$ satisfying
\be
i\partial_t W_{\pm}(t,t_1)
&=&
W_{\pm}(t,t_1)H_\pm(t)
\ee
reads
\be
H_\pm(t)
&=&
H_0
+
{\textstyle\int}_{\pm\infty}^{t} d^4x
\uu A_a(x){\textstyle\frac{\partial}{\partial x_0}} J^a(x).
\ee
In the next section we show the explicit form of $H_\pm(t)$
for a pointlike charge. As we shall see the Hamiltonian
has a clear physical interpretation.

\section{Explicit form of the new Hamiltonian for a pointlike charge}

A pointlike charge $q$ localized on an infinitely long world-line
$z^a(t)=\big(t,\bm z(t)\big)$ leads to the conserved current \cite{KC,CK}
\be
J^a(t,\bm x)
&=&
q\big(1,{\bm v}(t)\big)
\delta^{(3)}\big(\bm x -\bm z(t)\big)
\ee
where ${\bm v}(t)=d\bm z(t)/dt$. Let us assume that the world-line represents a charge which is at rest for times $t<t_-$ and $t>t_+$.
The assumption implies also that $\bm v(t_\pm)=0$ if we assume that
$t\mapsto\bm z(t)$ is twice differentiable. Under these assumptions we find
for $t_-\leq t\leq t_+$ that
\be
H_\pm
&=&
H_0
-
q\uu {\bm A}\big(z(t)\big)\cdot\bm v(t)
-
q{\textstyle\int}_{z(t_\pm)}^{z(t)}
d\bm z\cdot\uu {\bm E}.\label{h(z)}\\
H_-
&=&
H_0 \quad{\rm for}\quad t\leq t_-\\
H_+
&=&
H_0 \quad{\rm for}\quad t\geq t_+
\ee
The line integral in the third term of (\ref{h(z)}) is along the part of the charge world-line where the charge velocity is nonzero. The electric field operator takes the usual form
$\uu{\bm E}=-\partial_0 \uu{\bm A}-\bm\nabla \uu{A}{_0}$.

Let us note that the terms explicitly involving $\uu{A}{_0}$ have cancelled out.
It is clear from the construction that the electric field occuring in $H_\pm$ is free.
Therefore, the Hamiltonian does not contain self-energy terms but, instead, takes into account the work performed by the particle against the electric field.

\section{Photon statistics}

The operator $U(t,\pm\infty)$ (as well as $U_0(t,\pm\infty)$)
in general does not exist due to the problem with divergent phase factor.
Fortunately we do not really need $U(t,t_0)$ itself, but
only its action on operators $X$
\be
\bm U(t,t_0)(X)
=
U(t,t_0)^{\dag}X\, U(t,t_0)
\ee
Similarly, in order to compute the $S$-matrix we concentrate on
the limiting operator map
$\bm S=\bm U(+\infty,-\infty)$.

Eqs.~(\ref{uu A^H}), (\ref{Aret}), (\ref{Aadv}) imply that at one hand
\be
\uu A_a^{\rm ret}(x)
&=&
\lim_{\tau_0\to -\infty}\lim_{\tau\to +\infty}
U(\tau,\tau_0)^{\dag}
\uu A_a^{\rm adv}(x)
U(\tau,\tau_0)
=
\bm S\big(\uu A_a^{\rm adv}(x)\big)
\ee
and on the other
\be
\uu A_a(x)+{\textstyle\int} d^4y \uu D(x-y)J_a(y)
&=&
\bm S\big(\uu A_a(x)\big).
\ee
Finally, as shown in \cite{II}, the $S$-matrix $\bm S$
gives the action of the displacement operator on the field operators.

More interesting is the question of photon statistics in fields
produced by classical currents, especially if the currents are
pointlike or stationary. Accelerated pointlike charges lead, in
the standard formalism, to infrared catastrophe.
In a naive approach, all transition probabilities
are zero, which contradicts unitarity of the $S$-matrix. In
manifestly covariant approaches, such as the Gupta-Bleuler
formulation, static charge distributions lead to infinite
vacuum-to-vacuum probability, which again makes no sense.
A mathematically correct treatment is possible \cite {ND04},
but requires a change to an abstract representation. This goes against
the philosophy of explicit construction, adhered to in the present
paper.

In our approach infrared divergence can be avoided by requiring
that the vacuum probability density $Z(\bm k)$ tends to zero
at ${\bm k}=0$. Let us take
the operator $H_2(t)$ given by (\ref{H_2}) and split it into the
parts $H_2^{(30)}(t)$ and $H_2^{(12)}(t)$ involving, respectively,
the fields of spin-0 (i.e. $\uu a(\bm k,3)$ and $\uu a(\bm k,0)$) and spin-1
(i.e. $\uu a(\bm k,1)$ and $\uu a(\bm k,2)$).
Let us recall that the $N$-oscillator Hilbert space is spanned by
$N$-fold tensor products of vectors of the form
$|\bm k,n_1,n_2,n_3,n_0\rangle$. We shall refer to such
1-oscillator states as containing $n_1+n_2$ transverse
excitations and $n_3+n_0$ longitudinal ones.
Any state belonging to the $N$-oscillator Hilbert space
$\uu {\cal H}$ and containing $n$ transverse excitations,
where $n$ is is the sum of the transverse excitations of
all the $N$ oscillators, is regarded as a state involving
$n$ transverse photons. Similarly we define a general
state involving $n'$ longitudinal photons. In particular, the vector
\be
H_2^{(12)}(t_1)^{\dag}\dots H_2^{(12)}(t_n)^{\dag}|\uu O\rangle
\ee
belongs to the subspace of $n$-transverse-photon states. The state
\be
H_2^{(30)}(t_1)^{\dag}\dots H_2^{(30)}(t_{n'})^{\dag}|\uu O\rangle
\ee
involves $n'$ longitudinal photons.

Denote by ${\cal P}_{nn'}$ the projector on the subspace of
$\uu {\cal H}$ that contains states with $n$ transverse and
$n'$ longitudinal photons. The probability of
finding $n$ transverse and $n'$ longitudinal photons in the
state produced from vacuum by a classical current is thus
\be
p_{nn'}(t,t_0)
&=&
\langle \uu O|U(t,t_0)^{\dag}{\cal P}_{nn'}U(t,t_0)|\uu O\rangle.
\ee
Employing (\ref{cbh2}) we find for $p_{nn'}=p_{nn'}(\infty,-\infty)$
\be
p_{nn'}
&=&
\frac{1}{n!n'!}
\langle \uu O|
F_{12}^{n}e^{-F_{12}}
F_{30}^{n'}e^{-F_{30}}
|\uu O\rangle
=
\frac{1}{n!n'!}
\frac{d^{n}}{d\mu^{n}}
\frac{d^{n'}}{d\nu^{n'}}
C(\mu,\nu)\Big|_{\mu=\nu=-1}
\nonumber\\
C(\mu,\nu)
&=&
\langle \uu O|
e^{\mu F_{12}}
e^{\nu F_{30}}
|\uu O\rangle
\nonumber\\
F_{12}
&=&
\textstyle{\int_{-\infty}^{\infty}}d\tau
\textstyle{\int_{-\infty}^{\infty}}d\tau'
[ H_2^{(12)}(\tau),H_2^{(12)}(\tau')^{\dag}]
\nonumber\\
F_{30}
&=&
\textstyle{\int_{-\infty}^{\infty}}d\tau
\textstyle{\int_{-\infty}^{\infty}}d\tau'
[ H_2^{(30)}(\tau),H_2^{(30)}(\tau')^{\dag}]
\nonumber
\ee
Assuming that detectors react only to spin-1 photons we obtain photon statistics
\be
p_n=\sum_{n'=0}^\infty p_{nn'}
=
\frac{1}{n!}
\langle \uu O|
F_{12}^{n}e^{-F_{12}}
|\uu O\rangle
\ee
Alternatively, in order to describe quantum optics of spin-1
observables directly, without any reference to the spin-0
fields, we can consider states defined via reduced density
matrices with the spin-0 parts traced out. Probability $p_n$
is an example of an average computed in terms of such a reduced density matrix.

\subsection{Fourier description}

To find an explicit formula we first consider a current whose
4-dimensional Fourier transform is a well behaved function
$\tilde J_a(k)=\int d^4x J_{a}(x)e^{ik\cdot x}$. Then
\be
F_{12}
&=&
\int d\Gamma(\bm k) \uu I(\bm k)
\Big(
|\tilde J^a(\bm k)m_{a}(\bm k)|^2
+
|\tilde J^a(\bm k)\bar m_{a}(\bm k)|^2
\Big)
\nonumber
\\
F_{30}
&=&
\int d\Gamma(\bm k) \uu I(\bm k)
|\tilde J^a(\bm k)\omega_{a}(\bm k)|^2
\nonumber
\ee
where $\tilde J^a(\bm k)$ is a restriction of $\tilde J_a(k)$ to the light-cone.
Due to the continuity equation $\tilde J^a(\bm k)k_a(\bm k)=0$ the spin-1 expression
$F_{12}$ is independent of the choice of $\omega_A(\bm k)$.

Employing the relation between $\uu I(\bm k)$ and
$I(\bm k)=|\bm k\rangle\langle\bm k|\otimes 1$ we can write the generating function as
\begin{widetext}
\be
C(\mu,\nu)
&=&
\Bigg({\textstyle\int} d\Gamma(\bm k) Z(\bm k)
\exp\Big(
\frac{\mu}{N}
\big(
|\tilde J^a(\bm k)m_{a}(\bm k)|^2
+
|\tilde J^a(\bm k)\bar m_{a}(\bm k)|^2
\big)
\Big)
\exp\Big(
\frac{\nu}{N}
|\tilde J^a(\bm k)\omega_{a}(\bm k)|^2
\Big)
\Bigg)^N
\ee
Of particular interest is the thermodynamic limit $N\to\infty$ for the spin-1 part.
The corresponding generating function becomes
\be
\lim_{N\to\infty}C(\mu,0)
=
\exp\Big(
\mu
{\textstyle\int} d\Gamma(\bm k) Z(\bm k)
\big(
|\tilde J^a(\bm k)m_{a}(\bm k)|^2
+
|\tilde J^a(\bm k)\bar m_{a}(\bm k)|^2
\big)
\Big)
\ee
and
\be
p_n
&=&
{\textstyle\frac{1}{n!}}\Big({\textstyle\int} d\Gamma(\bm k) Z(\bm k)
\big(
|\tilde J^a(\bm k)m_{a}(\bm k)|^2
+
|\tilde J^a(\bm k)\bar m_{a}(\bm k)|^2
\big)
\Big)^n
\exp\Big(
-
{\textstyle\int} d\Gamma(\bm k) Z(\bm k)
\big(
|\tilde J^a(\bm k)m_{a}(\bm k)|^2
+
|\tilde J^a(\bm k)\bar m_{a}(\bm k)|^2
\big)
\Big)\nonumber
\ee
\end{widetext}
This is basically the well known Poisson distribution,
with one modification: The standard infrared-divergent
result is found if one puts $Z(\bm k)=1$. However, we
know that
${\textstyle\int} d\Gamma(\bm k) Z(\bm k)=1$ and thus
$Z(\bm k)\neq 1$. We have remarked earlier that the
maximum value of $Z(\bm k)$ is a positive Poincar\'e
invariant, denoted by $Z$.
Introducing the new function $\chi(\bm k)=Z(\bm k)/Z$,
and absorbing $Z^{1/2}$ into a renormalized current
$\tilde J^a_{\rm ren}(\bm k)=Z^{1/2}\tilde J^a(\bm k)$ we find
\begin{widetext}
\be
p_n
&=&
{\textstyle\frac{1}{n!}}
\Big({\textstyle\int} d\Gamma(\bm k) \chi(\bm k)
\big(
|\tilde J^a_{\rm ren}m_{a}(\bm k)|^2
+
|\tilde J^a_{\rm ren}\bar m_{a}(\bm k)|^2
\big)
\Big)^n\exp\Big(
-
{\textstyle\int} d\Gamma(\bm k) \chi(\bm k)
\big(
|\tilde J^a_{\rm ren}m_{a}(\bm k)|^2
+
|\tilde J^a_{\rm ren}\bar m_{a}(\bm k)|^2
\big)
\Big)\nonumber
\ee
\end{widetext}
Now this is indeed the standard {\it regularized\/} expression.
The latter provides us with a new information about the vacuum
wave function $O(\bm k)$: It has to vanish at the origin
$\bm k=\bm 0$ if one wants the cut-off function $\chi(\bm k)$ to
regularize the infrared divergence. The origin belongs to the boundary of the light cone. Vanishing at the origin is a Poincar\'e invariant boundary condition.

We regard this result as very important, as it handles in a natural manner two elements
that are imposed in an ad hoc manner in the standard formalism.
First of all, we do not need to justify the infrared cut-off by
hand-waving arguments on unobservability of ``soft photons". Our
formalism introduces the cutting-off function automatically.
Secondly, we know what is the origin of a renormalization constant:
This is simply the Poincar\'e invariant associated with the vacuum wave function.

\subsection{Pointlike static charge}

In this case it makes no sense to switch to the Fourier domain,
since the position space calculation is more straightforward.
Assume the current is $J_a(x)=(q\delta^{(3)}(\bm x),\bm 0)$.
The generating function becomes
\begin{widetext}
\be
C(\mu,\nu)
&=&
\Big({\textstyle\int} d\Gamma(\bm k) Z(\bm k)
e^{\textstyle
(2q^2/|\bm k|^2)
\big(
\frac{\mu}{N}
x_{0}(\bm k)^2
+
\frac{\mu}{N}
y_{0}(\bm k)^2
+
\frac{\nu}{N}
z_{0}(\bm k)^2
+
\frac{\nu}{N}
t_{0}(\bm k)^2
\big)}
\Big)^N\label{154}
\ee
In the thermodynamic limit
\be
\lim_{N\to\infty}
C(\mu,\nu)
&=&
\exp\Big({{\textstyle\int} d\Gamma(\bm k) Z(\bm k)
(2q^2/|\bm k|^2)
\big(
\mu
x_{0}(\bm k)^2
+
\mu
y_{0}(\bm k)^2
+
\nu
z_{0}(\bm k)^2
+
\nu
t_{0}(\bm k)^2
\big)}\Big)\label{155}
\ee
\end{widetext}
Identical results are obtained if instead of $U(\infty,-\infty)$ one works with
$U(t,\pm\infty)$ for a finite $t$.

Let us remark that an analogous calculation performed in a reducible
version of Gupta-Bleuler formalism \cite{MC-GB} leads to
vacuum-to-vacuum ``probabilities"
that are greater than 1. The reason is that for currents whose only nonzero component is
$J_0(x)$ the Fourier-space version of continuity equation does not read
$\tilde J_a(k)k^a=0$, but $\tilde J_0(k_0)\delta(k_0)=0$, and one cannot claim that
$\tilde J_a(k)$ is spacelike.
In our formalism the timelike component of the current comes with the correct sign.

\subsection{R\'enyi statistics for finite $N$}

Generating functions can be written in a unified way for any $N$ in
terms of Kolmogorov-Nagumo averages of the form used in R\'enyi statistics.
Let us recall that R\'enyi's alpha entropies were obtained in
\cite{Renyi} as Kolmogorov-Nagumo averages
\be
\langle I\rangle_\phi=\phi^{-1}\big(\sum_j p_j \phi(I_j)\big)
\ee
of the random variable $I_j=\ln(1/p_j)$, and $\phi(x)=e^{(1-\alpha)x}$.
For $\alpha=1$ one obtains the standard Boltzmann-Shannon entropy.
In \cite{CN} it was shown that thermodynamics that employs R\'enyi
type averaging can be used to derive certain equilibrium distributions
occuring in linguistics and protein folding. Various arguments based
on thermodynamics suggest that $\alpha\neq 1$ statistics may be
typical of finite systems.
Photon statistics for finite-$N$ representations of CCR supports
this intuition.

Indeed, in the thermodynamic limit we found a generating function
of the form
$
C(\mu,0)
=e^{\langle j(\mu)\rangle}
$
with
\be
\langle j(\mu)\rangle
=
\mu
{\textstyle\int} d\Gamma(\bm k) Z(\bm k)
\big(
|\tilde J^a(\bm k)m_{a}(\bm k)|^2
+
|\tilde J^a(\bm k)\bar m_{a}(\bm k)|^2
\big)\nonumber
\ee
being a linear ($\alpha=1$) average of $\mu
\big(
|\tilde J^a(\bm k)m_{a}(\bm k)|^2
+
|\tilde J^a(\bm k)\bar m_{a}(\bm k)|^2
\big)
$,
with probability density $Z(\bm k)$. For finite $N$ we find
$
C(\mu,0)
=e^{\langle j(\mu)\rangle_\phi}
$
where $\phi(x)=e^{(1-\alpha)x}$, $\alpha=1-1/N$, and we average
the same random variable with the same probability distribution.
Obviously, the limits $N\to\infty$ and $\alpha\to 1$ are equivalent.
It follows that the field theories based on $N<\infty$ or
$N=\infty$ representations are related to one another in a
way that is analogous to the relation between systems described
by $0<\alpha<1$ and $\alpha=1$ entropies. These, on the other hand,
are known to apply to fractal and non-fractal geometries, respectively.
A natural intuition thus relates the $N<\infty$ case to some
``space-time foam", and $N=\infty$ to continuum
space-time.

\section{Classical fields produced by classical sources:
A quantum way}\label{Qway}

Our previous analysis shows that, having a classical current $J_a(x)$,
we obtain a result that agrees with standard calculations if one
(a) absorbs $Z^{1/2}$ in the current by means of
$J_a^{\rm ren}(x)=Z^{1/2}J_a(x)$ (bare charge renormalization
$q\mapsto q^{\rm ren}=Z^{1/2}q$), and (b) compares the result with
a regularized formula, which is anyway the one we have to compare with experiment.
$Z$ is not a constant but rather an invariant of the
Poincar\'e group that characterizes a given vacuum. We have also obtained a cut-off function
$\chi(\bm k)=Z(\bm k)/Z$. At this stage we do not have much
information as to the exact form of $\chi(\bm k)$ and can
only say that it vanishes
for large $\bm k$ and $\bm k=0$,  and that
$\chi(\bm k)/|\bm k|$ fulfills the assumptions of the Riemann-Lebesgue lemma.

Now let us take an arbitrary classical amplitude $\alpha(\bm k,\pm)$
corresponding to left- and right-handed Fourier modes of a classical
electromagnetic field. We define the {\it quantum optics regime\/}
by the support of those classical amplitudes that satisfy
\be
\alpha(\bm k,\pm)
&=&
\alpha(\bm k,\pm)
\chi(\bm k).\label{qor}
\ee
The latter formula is meaningful provided $\chi(\bm k)=1$ if
$\bm k$ belongs to quantum optics regime.
Classical fields belonging to quantum optics regime do not
contain wavelenghts that are either too large or too small.
Let $|\uu O(\alpha)\rangle$ be a coherent state with $\alpha$
in quantum optics regime. Let us take the coherent state
average of the retarded solution of Heisenberg equation of
motion (\ref{Aret}) and express it in terms of the renormalized current
$J_a^{\rm ren}(x)$. Taking into account that
\begin{widetext}
\be
{\cal A}_a(x)
&=&
\langle\uu O(\alpha)|{\uu A}_a(x) |\uu O(\alpha)\rangle
=
Zi{\textstyle\int} d\Gamma(\bm k)
\big(
m_{a}(\bm k)
\alpha(\bm k,+)
+
\bar m_{a}(\bm k)
\alpha(\bm k,-)
\big)e^{-ik\cdot x}
+ {\,\rm c.c.},
\ee
\end{widetext}
by (\ref{<A>}) and the assumption that (\ref{qor}) is fulfilled, the formula
\be
\langle\uu O(\alpha)|\uu D(x-y)|\uu O(\alpha)\rangle
 =
Zi{\textstyle\int} d\Gamma(\bm k)
\chi(\bm k)
\big(
e^{-ik\cdot (x-y)}
-
e^{ik\cdot (x-y)}
\big)
\label{class J-P},
\ee
and dividing the entire solution by $Z^{1/2}$,
we find that the classical field
\be
\langle A_a^{\rm ret}(x)\rangle
&=&
Z^{-1/2}{\cal A}_a(x)
+
\textstyle{{\textstyle\int}} d^4y D_{\rm ret}(x-y)J_a^{\rm phys}(y)
\label{eff A}\\
\partial^a\langle A_a^{\rm ret}(x)\rangle
&=&
\partial^a{\cal A}_a(x)=
0\\
\partial^a J_a^{\rm phys}(x)
&=&0
\ee
exhibits the  textbook relation between the in-field,
renormalized current, and renormalization constant.

Here $D_{\rm ret}(x-y)$ is the ordinary retarded Green
function and
$J_a^{\rm phys}(y)$ is the efective current obtained after
charge renormalization and inclusion of $\chi(\bm k)$ in its Fourier transform (the convolution of
$\uu D_{\rm ret}(x-y)$ and
$J_a(y)$ in (\ref{Aret}) allows to shift the regularization
from the Green function to the current, and vice versa).
All these objects have occured in our calculation automatically.

Finally, let us have a closer look at the effective current.
For simplicity take a static pointlike charge.
Employing the relations
\be
Z^{1/2}J_0^{\rm phys}(x_0,\bm x)
&=&
\int d^4y \langle \uu O|\uu \delta(x-y)|\uu O\rangle J_0(y)
=
q\langle \uu O|\partial_0\uu D(0,\bm x)|\uu O\rangle
=
Zq\textstyle{\frac{1}{2}\int \frac{d^3k}{(2\pi)^3}}\chi(\bm k)
\big(e^{i\vec k\cdot \vec x}+e^{-i\vec k\cdot \vec x}\big)
\nonumber
\ee
we find that the effective total charge
\be
Q&=&
\int d^3xJ_0^{\rm phys}(x_0,\bm x)
=
Z^{1/2}q\chi(\bm 0)=q^{\rm ren}\chi(\bm 0)
\nonumber
\ee
vanishes since we require $\chi(\bm 0)=0$.

In order to check the physical meaning of this condition
let us consider the simple case of spherically symmetric
\be
\chi_{k_1,k_2}(\bm k)
&=&
\left\{
\begin{array}{ll}
1 & {\rm for}\, k_1\leq|\bm k|\leq k_2\\
0 & {\rm otherwise}
\end{array}
\right.
\ee
Then $Q=q^{\rm ren}$ for $k_1=0$ and $Q=0$ for $k_1>0$.
One can check that the effective charge density
$J_0^{\rm phys}(x_0,\bm x)=\rho_{k_2}(\bm x)-\rho_{k_1}(\bm x)$, where
$\int d^3x \rho_{k_2}(\bm x)=\int d^3x \rho_{k_1}(\bm x)=q^{\rm ren}$
for all $k_2,k_1>0$, but simultaneously the pointlike limit
$\lim_{k_1\to 0}\rho_{k_1}(\bm x)=0$ holds.
It turns out that the charge density consists of a difference
of two densities: One, which is the sharper and more localized
the greater $k_2$, and the other which is the flatter and
less localized the smaller $k_1$. One of them corresponds
to localization of the charge
$q^{\rm ren}$ in a sinc-like very sharp-peaked density,
and the other describes
the charge $-q^{\rm ren}$ distributed almost uniformly
in a volume which becomes infinite if
$k_1=0$. Plots of the densities for $k_1>0$ and $k_1=0$
become indistinguishable even for relatively large $k_1$
and small $k_2$, so we leave this exercise to the readers.
In practice, one cannot locally distinguish between
$k_1=0$ and $k_1\approx 0$, but globally the two cases are inequivalent.

\section{Summary and conclusions}

We have discussed a new quantization scheme based on reducible
representations of CCR. The principal goal of this research
program is to arrive at a mathematically consistent formalism
for quantum fields, that should be described by a {\it theory\/}
and not a {\it set of working rules\/}, as Dirac summarized the
current status of field quantization \cite{Dirac1984a}. In our
approach fields are represented by operators and not
operator-valued distributions. The field is a finite quantum system, and
the measure of its size is the parameter $N$. For this reason
two sources of infinities are absent in the formalism from the
very outset: We can multiply field operators at the same points
in space-time and all the tensor products one encounters are finite.
The latter condition means that we deal only with factors of type I,
in von Neumann's terminology.

We have carefully analyzed Poincar\'e covariance of the theory.
There were two aspects we had to understand to make sure that
the new quantization is not inconsistent with special relativity.
First of all, we constructed a unitary representation of the
Poincar\'e group whose carrier space is an ordinary Hilbert
space involving no indefinite metric. Four-potential operator
is self-adjoint, the dynamics is unitary, but commutation
relations for fields are nevertheless manifestly Poincar\'e
covariant. The formalism is a promising alterantive to the
Gupta-Bleuler quantization, where the price payed for manifest
covariance is either in non-positivity of the Hilbert-space metric,
or in non-Hermicity of the potential.

Secondly, we had to understand in what way a Lorentz transformation
influences the gauge freedom. The latter has led to the observation
that a change of gauge due to Lorentz transformations can be always
compensated by a Bogoliubov unitary transformation of the vacuum.
An inclusion of the Bogoliubov transformation turns the 4-potential
into a 4-vector field.

One element that remained arbitrary is what kind of a tetrad one has
to associate with a 4-potential. Once one makes
a choice then the
remaining freedom can be controlled by Bogoliubov transformations, whose explicit form has been given.
In our formalism the dynamics is unitary, in the ordinary
meaning of this word. We have no problems with negative or greater than 1 ``probabilities" that occur in the Gupta-Bleuler formalism. The correct probability interpretation
is guaranteed by the Schwartz inequality.
Further, if one looks at the radiation fields then our formalism
produces the standard regularized formula, which does not depend
on a choice of gauge.

Finally, its seems that we have produced the first example of
Heisenberg dynamics where the retarded or advanced fields
unitarily evolve from, say, $t=0$ to another finite $t$. Our
construction allowed to systematically derive the form of
Hamiltonian that is responsible for such an evolution, and the
result turned out to differ from the usual minimal-coupling expression:
There is no scalar-potential part and a new term occurs. The term
describes the work performed by the charge moving in electric field. This result may have implications for quantum optics where the usual treatments of spontaneous emission or resonance fluorescence are based on initial conditions at $t=0$ and not $t=-\infty$
(cf. \cite{Mandel,Rz1}, the exception is \cite{Pachucki}).

Summing up, we think we have proposed at least a nontrivial answer
to the problem posed by Dirac in his last two papers.
It looks like the formalism, supplemented by its fermionic analogue
introduced in \cite{III}, is ready for calculations in full quantum
electrodynamics. Some preliminary results on loop diagrams have
been already obtained and will be reported in a future paper.

\acknowledgments

We are indebted to Professor David Finkelstein and Ms. Klaudia
Wrzask for discussions. This work was performed as a part of the
bilateral Flemish-Polish project ``Soliton techniques applied to
equations of quantum field theory". The research of M.C. was also
supported by the Polish Ministry of Scentific Research and
Information Technology under the (solicited) grant No.
PBZ-Min-008/P03/2003.


\begin{thebibliography}{99}
\bibitem{Klaudia}This explict construction has been done in collaboration with K.~Wrzask.
\end{thebibliography}
\end{document}